\documentclass[aps,pra,reprint,superscriptaddress]{revtex4-2}

\usepackage[utf8]{inputenc}
\usepackage{amsthm}
\usepackage{amsmath}
\usepackage{xfrac}
\usepackage{color}
\usepackage{hyperref}
\usepackage{graphicx}
\usepackage[dvipsnames]{xcolor}
\usepackage{amssymb}
\usepackage{bbold}
\usepackage{gensymb}
\usepackage{afterpage}
\usepackage[normalem]{ulem}
\usepackage{tikz}

\newcommand{\id}{\mathbb{1}} 

\def\iu{\ensuremath{\mathrm{i}}}
\def\i{\iu}

\renewcommand{\d}{\mathrm{d}}

\newcommand{\ket}[1]{|#1\rangle} 
\newcommand{\ketbra}[2]{|#1\rangle\!\langle #2|} 
\newcommand{\tr}{\mathrm{tr}} 

\newcommand{\Dyn}{\mathcal{D}}
\newcommand{\cpt}{\mathcal{E}}

\newcommand{\dynamics}[1]{(#1)}

\newcommand{\Sys}{\mathrm{S}} 
\newcommand{\Anc}{\mathrm{A}} 
\newcommand{\dtr}{d_\mathrm{tr}}
\newcommand{\mon}{C}
\newcommand{\esq}{\mon_\mathrm{sq}}

\theoremstyle{plain}
\providecommand{\definitionname}{Definition}

\newcommand{\Env}{\textrm{E}}

\newcommand{\F}{\textrm{F}}
\newcommand{\SA}{{\Sys\!\Anc}}
\newcommand{\rhoSA}{\rho_\SA}

\newcommand{\rot}[1]{\rotatebox{90}{#1}}
\newcommand{\cm}{\checkmark}

\usetikzlibrary{patterns}
\usetikzlibrary{patterns.meta}
\definecolor{mathematicablue}{rgb}{0.87,0.94,1}
\definecolor{mathematicablue}{rgb}{0.81,0.88,1}
\definecolor{mathematicamediumblue}{rgb}{0.56,0.71,1}
\definecolor{mathematicadarkblue}{rgb}{0.368417, 0.506779, 0.709798}
\definecolor{mathematicaorange}{rgb}{1,0.9,0.8}
\definecolor{mathematicadarkorange}{rgb}{1,0.5,0}
\definecolor{mathematicagreen}{rgb}{0.56,0.69,0.19}

\usepackage{tikz}
\usetikzlibrary {shadows.blur, decorations.pathreplacing, arrows, positioning} 

\newcommand{\qwire}[3]{
	\draw[line width=1.2pt, blur shadow={shadow blur steps=5, shadow xshift=0.3mm, shadow yshift=-0.3mm}] (#1, #2) -- (#1+#3, #2);
}

\newcommand{\qbox}[4]{
	\filldraw[line width=1.2pt, rounded corners, fill=Orchid!20!white,  draw=Plum!80!black, blur shadow={shadow blur steps=5, shadow xshift=0.3mm, shadow yshift=-0.3mm}] (#1, -#2-1.4) rectangle ++(#4,1.5) node[midway]{#3};
}

\newcommand{\qcirc}[3]{
	\filldraw[line width=1.2pt, rounded corners, fill=white,  draw=white] (#1, #2) circle (0.3cm) node {#3};
}

\newcommand{\centerbox}[3]{
    \qwire{#1+0.7}{#2-1.1}{0.3}
    \qwire{#1+0.7}{#2-0.2}{0.3}
    \qbox{#1}{0}{#3}{0.7}
}

\newcommand{\unitarybox}[5]{
    \qcirc{#1-0.6}{#2-0.2}{#4}
    \qcirc{#1-0.6}{#2-1.1}{#5}
    \qwire{#1}{#2-1.1}{-0.3}
    \qwire{#1+0.7}{#2-1.1}{0.3}
    \qwire{#1}{#2-0.2}{-0.3}
    \qwire{#1+0.7}{#2-0.2}{0.3}
    \qbox{#1}{0}{#3}{0.7}
}

\newcommand{\traceout}[2]{
\draw[line width=1.2pt, blur shadow={shadow blur steps=5, shadow xshift=0.3mm, shadow yshift=-0.3mm}] (#1-0.15, #2-0.2) -- (#1+0.1, #2+0.1);
\draw[line width=1.2pt, blur shadow={shadow blur steps=5, shadow xshift=0.3mm, shadow yshift=-0.3mm}] (#1-0.05, #2-0.2) -- (#1+0.2, #2+0.1);
}

\begin{document}

\author{Charlotte Bäcker}
\affiliation{Institut f{\"u}r Theoretische Physik, Technische Universit{\"a}t Dresden, 
D-01062, Dresden, Germany}

\author{Nick Maryshchak}
\affiliation{Institut f{\"u}r Theoretische Physik, Technische Universit{\"a}t Dresden, 
D-01062, Dresden, Germany}

\author{Walter T. Strunz}
\affiliation{Institut f{\"u}r Theoretische Physik, Technische Universit{\"a}t Dresden, 
D-01062, Dresden, Germany}

\title{How rare are Markovian quantum dynamics?}

\begin{abstract}
A profound understanding of decoherence and dissipation in quantum dynamics is crucial for the realistic modeling of the evolution of quantum systems. 
In open quantum dynamics one distinguishes between a memoryless, so-called Markovian evolution and dynamics incorporating memory effects, termed non-Markovian. In this work we study how prevalent memory effects are in the set of all such dynamics. We thus investigate how often a Markovian description is applicable. This question is approached by investigating randomly generated two-step qubit dynamics with respect to different concepts and witnesses of non-Markovianity. 
We observe that almost all dynamics are non-Markovian, and only a small (yet finite) fraction is Markovian. Furthermore, we study how this proportion changes when considering certain subclasses such as lower rank or mixed-unitary dynamics.
Importantly, our results shed light on the relative ratios of -- and interrelations between -- the sets of dynamics that are non-Markovian with respect to different criteria.
Finally, we investigate the fraction of dynamics in which the memory effects are necessarily of quantum nature and establish a connection between two recently developed concepts that characterize the quantumness of memory in non-Markovian dynamics.
\end{abstract}\maketitle

\section{Introduction}

A general aim of physics is to describe a phenomenon as exactly as necessary, but at the same time as simply as possible. 
For the dynamics of quantum systems the first and thus also best-known approach for a theoretical description is given in terms of the Schrödinger equation.
The validity of the Schrödinger equation is based on the crucial assumption that the system is isolated. In realistic dynamics, however, quantum systems tend to interact with their environment and thus the resulting irreversible effects have to be incorporated in the theoretical modelling. Hence, the formalism of quantum dynamics can be extended in order to describe systems under external, irreversible influence.
A widely used approach based on the work by Gorini, Kossakowski, Sudarshan and Lindblad formalized this idea mathematically in terms of the so-called GKSL or Lindblad master equation~\cite{GorKosSud1976, Lin1976}. Due to the semigroup property of this description it is often referred to as \emph{Markovian quantum dynamics}. This is due to the fact that the dynamical map reflects the sequential transition appearing in the classical Chapman-Kolmogorov equation in terms of the transition probability of a classical Markov process.
Clearly, the GKSL-approach has proven to be a very powerful tool in many areas of open quantum dynamics.
Nevertheless, it describes the irreversible loss of information or energy into the environment under the assumption that temporal correlations within the evironment decay instantanously.
Thus, the GKSL equation does not reflect a generic system-environment situation, where the system affects its environment at some early time such that later, the environment acts back onto the system in a way that is correlated to this earlier interaction.
Such quantum dynamics is said to show memory-effects or to be \emph{non-Markovian}. Then, the future evolution will not exclusively depend on the current state, but it is affected by degrees of freedom correlated with the earlier dynamics (be they classical or quantum). In other words, it is the full history of the dynamics that determines the future evolution \cite{VacSmiLaiPiiBre2011}.

In more recent years not only the presence but also the nature of memory effects has been a subject of intense research. Within several frameworks it was shown that although the non-Markovian dynamics itself is quantum, memory effects need not be of quantum nature \cite{MilEglTarThePleSmiHue2020, BaeBeyStr2024, GiaCos2021, TarQuiMurMil2023, BusGanGosBadPanMohDasBer2025}.

In order to describe quantum dynamics with memory effects,
several conceptual \cite{BreLaiPii2009, RivHuePle2010, HalCreLiAnd2014, liConceptsQuantumNonMarkovianity2018, pollockNonMarkovianQuantumProcesses2018} and numerical \cite{HOPS2014, tanimuraNumericallyExactApproach2020, cygorekSimulationOpenQuantum2022, LinTuStr2024TensorNetworks} approaches have been developed.
While the Markovian GKSL master equation is a well-understood and computationally straightforward way to treat quantum dynamics, all approaches to quantum non-Markovianity have in common that they are much more complex and require new concepts and numerical tools. 
Indeed, a Markovian treatment is often reasonable and most convenient. Here, however, we want to address the question whether a {\it generic} open quantum dynamics falls into the Markovian or non-Markovian class.
These considerations may help to gauge whether the quantum dynamics of interest really requires an involved non-Markovian treatment or whether a Markovian description is essentially sufficient.

As Markovianity in classical physics is based on joint probabilities distributed over time, there is no unique analogue approach to quantum non-Markovianity due to the fundamental influence of quantum measurements~\cite{VacSmiLaiPiiBre2011}. Thus, various definitions and criteria for non-Markovianity in the quantum realm have been proposed \cite{BreLaiPii2009, RivHuePle2010, RivHuePle2014, LuoFuSong2012DPI, lorenzoGeometricalCharacterizationNonMarkovianity2013a, LuWanSun2010, scandiQuantumFisherInformation2025, ChrKos2012}. They differ in their physical motivation, the necessary assumptions on the environment, their sensitivity and the complexity to compute them.

In this work we want to address the question how frequent non-Markovianity compared to Markovianity appears in generic quantum dynamics. By generic we mean among a random selection of dynamics, sampled according to a reasonable measure.
We aim at answering this question in the same spirit as previous research approached the question of the fraction of separable states in the set of bipartite states \cite{VolumeSeparableStates1998} or the question of how often randomly chosen preparations and measurements produce non-classical statistics \cite{rossiHowTypicalContextuality2025}.
A random matrix approach to Markovian open quantum dynamics (i.e. the GKSL equation) can be found in
Refs.~\cite{langeRandommatrixTheoryLindblad2021, chenRandomizedMethodSimulating2025, denisovUniversalSpectraRandom2019}.

Here, we will study the space of all quantum dynamics consisting of two time steps and investigate the fraction of non-Markovian quantum dynamics therein.
Given that there are multiple suitable definitions and criteria for non-Markovianity, we will compare how often each of them witnesses non-Markovianity and the quantumness of the memory effects. Moreover, our analysis will shed light on the interrelations (and sizes) of the sets of non-Markovian dynamics according to those different criteria.

Our paper is structured as follows. In Sec.~\ref{sec:theory-non-Markovianity} we review  concepts of quantum non-Markovianity and witnesses of memory effects, and we present the sampling method. Details and results of these investigations will be shown and discussed in Sec.~\ref{sec:results}, first in general and then for certain subclasses. In Sec.~\ref{sec:quantum-vs-classical} we first review concepts to distinguish between classical and quantum memory in non-Markovian dynamics and provide a statistical analysis of their occurrence. We close with a conclusion and outlook in Sec.~\ref{sec:conclusion}.

\section{Concepts and Terminology}
\label{sec:theory-non-Markovianity}
Quantum dynamics of a state can be captured by a completely positive and trace preserving (CPT) dynamical map $\cpt_t$ such that $\rho(t) = \cpt_t[\rho(0)]$ for $t\ge 0$, if the initial state is uncorrelated with its environment \cite{GorKosSud1976,Lin1976}. In this paper, we will mainly focus on dynamics involving only two discrete moments $0\leq t_1 \leq t_2$ of that continuous time. The two corresponding CPT maps are combined as a disrecte (two-time)
\emph{dynamics} $\Dyn=\dynamics{\cpt_1, \cpt_2} := \dynamics{\cpt_{t_1}, \cpt_{t_2}}$. Our investigations evolve around the question whether $\Dyn$ belongs to the class of non-Markovian dynamics and how reliably non-Markovianity can be detected with suitable witnesses.

Given $\Dyn$, for an invertible $\cpt_1$ we can define an intermediate map $\cpt_{21} = \cpt_{2} \circ \cpt_1^{-1}$ mapping the state $\rho(t_1)$ onto $\rho(t_2)$, 
such that
\begin{align}\label{eq:intermediate}
    \cpt_2 = \cpt_{21} \circ \cpt_1.
\end{align}
For all the $\Dyn$ we consider later, $\cpt_{21}$ was found to exist.

\subsection{Markovian Quantum Dynamics}
The prime example of Markovian quantum dynamics is provided by the GKSL master equation for the continuous evolution of open quantum systems (setting $\hbar=1$)~\cite{Lin1976, GorKosSud1976}
\begin{align}
    \label{eq:lindblad}
    \frac{\d\rho}{\d t} & = {\cal L} \rho \\ \nonumber
    & = - \i \left[H, \rho \right] + \sum_i \gamma_i\left[L_i \rho L_i^\dagger - \frac{1}{2}\left(L_i^\dagger L_i \rho + \rho L_i^\dagger L_i \right) \right],
\end{align}
where the generator ${\cal L}$ of time evolution contains the Hamiltonian $H$, decay rates $\gamma_i\geq0$ and so-called Lindblad or jump operators $L_i$. The corresponding dynamical map is thus given by the exponential
\begin{align}
    \label{eq:dynmaplindblad}
    \cpt_t = \exp{{\cal{L}} t},
\end{align}
which satisfies the semigroup property 
\begin{align}\label{eq:semigroup}
    \cpt_{t_2}=\cpt_{t_2-t_1}\circ\cpt_{t_1},\;\;\; t_2>t_1.
\end{align}
We see that in this case the intermediate map in Eq. \eqref{eq:intermediate} is given by 
\begin{align}
    \cpt_{21} = \exp{{\cal{L}} (t_2-t_1)},
\end{align}
and is thus itself a CPT map, too. 
Replacing the constant decay rates $\gamma_i$ in the generator ${\cal L}$ by positive, time dependent rates $\gamma_i(t)\ge 0$ we lose the simple exponential form \eqref{eq:dynmaplindblad} and thus the semigroup property \eqref{eq:semigroup} of the dynamical map. The intermediate map takes the form of a time-ordered exponential. Still, in this case too, $\cpt_{21}$ is a CPT map.

A general, given dynamics $\Dyn$, however, does not need to share this property: as we will study in this work, for generic CPT maps $\cpt_1$ and $\cpt_2$, the intermediate map $\cpt_{21}$ in \eqref{eq:intermediate} will not be a CPT map. This is due to the fact that correlations of the system with other degrees of freedom that exist at time $t_1$ become relevant for the ensuing dynamics from $t_1$ to $t_2$. We speak of memory effects
or {\it non-Markovian quantum dynamics}, a notion that is associated with various definitions and criteria.

\subsection{Non-Markovian Quantum Dynamics}
\label{sec:non-markovian-dynamics}

We will review some well-established definitions of non-Markovianity for two distinct times $t_1$ and $t_2$ and suitable corresponding criteria and witnesses. A more detailed overview can be found in Refs.~\cite{VacSmiLaiPiiBre2011, RivHuePle2014, HalCreLiAnd2014, liConceptsQuantumNonMarkovianity2018, breuerFoundationsMeasuresQuantum2012}.
Considering dynamics given by a master equation first, there are generalizations of the form \eqref{eq:lindblad} with, however, time-dependent rates
$\gamma_i(t)$. As mentioned earlier, for positive rates any intermediate map is still CPT, yet this ceases to be true for physically sound master equations with $\gamma_i(t)<0$ for certain $i$ and certain times $t$. 
Hence, as soon as at least one of the decay rates in \eqref{eq:lindblad} becomes negative, the dynamics can be called non-Markovian \cite{HalCreLiAnd2014}. 
For the time-discrete two-step dynamics $\Dyn=\dynamics{\cpt_1, \cpt_2}$ we are considering, a negative decay rate implies a non-CP intermediate map $\cpt_{21}$ for certain times $t_1,t_2$ which means $\Dyn$ is CP-indivisible \cite{RivHuePle2010}.
Thus, following general consensus, it suggests itself to call a two-step dynamics $\Dyn=(\cpt_1,\cpt_2)$ {\it non-Markovian}, if in \eqref{eq:intermediate}
the intermediate map $\cpt_{21}$
is not a completely positive (CP) map. This concept is closely related to the definition of Markovianity in classical probability theory based on transition probabilities \cite{RivHuePle2014}. In terms of physics it implies that for the intermediate dynamics from $t_1$ to $t_2$, correlations 
of the state $\rho(t_1)$ with its environment need to be taken into account when determining its ensuing dynamics to $t_2$: we encounter memory-effects.
Note that in contrast to the question of whether \emph{one} given CPT map is divisible, such as thoroughly treated in Refs.~\cite{wolfDividingQuantumChannels2008, DavZimPin2019, DavZim2023Divisibility}, we investigate a dynamics $\Dyn$ of \emph{two} maps. The question is thus not if there is at least one decomposition of $\cpt_2$ in terms of two arbitrary CPT maps, but if for  $\cpt_2$ and fixed first factor $\cpt_1$ there exist a suitable $\cpt_{21}$ which is CPT.

For a given $\Dyn$ non-Markovianity according to CP-indivisibility can directly be tested.
Checking whether a given dynamics is CP-indivisible is comparably simple, given that $\cpt_{1}$ is invertible. Computing $\cpt_{21} = \cpt_2 \circ \cpt_1^{-1}$
allows for a direct investigation of the intermediate map  $\cpt_{21}$ using the Choi-Jamiołkowski channel-state-duality \cite{choiCompletelyPositiveLinear1975, jamiolkowskiLinearTransformationsWhich1972}.
A quantum map $\cpt_\Sys$ on a system $\Sys$ is CP if and only if its corresponding Choi-Jamiołkowski state
\begin{align}
    \label{eq:def-choi}
    E = (\id_\Anc \otimes \cpt_\Sys) \Phi_{\Sys \Anc}^+
\end{align}
with $\Phi^+_{\Sys\Anc}=\ketbra{\Phi^+_{\Anc \Sys}}{\Phi^+_{\Anc \Sys}}$ being a maximally entangled state of system $\Sys$ and ancilla $\Anc$, is positive semidefinite. Thus, identifying the sign of the eigenvalues of $E_{21}$ reveals whether the dynamics $\Dyn=\dynamics{\cpt_1, \cpt_2}$ is CP-indivisible and hence non-Markovian according to this definition.

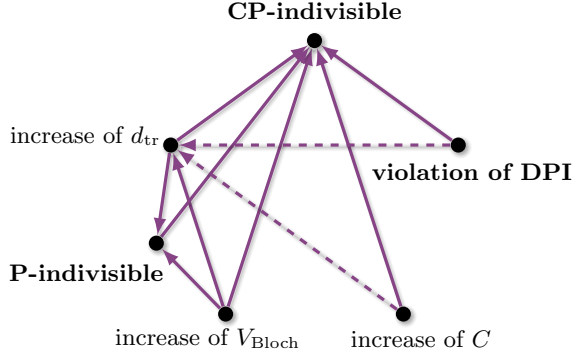
\begin{figure}
    \centering

  \begin{tikzpicture}[bullet/.style={circle, fill, inner sep=2pt}, implication/.style={-latex,Plum!80!black,line width=1.2pt, blur shadow={shadow blur steps=5, shadow xshift=0.3mm, shadow yshift=-0.3mm}}, conjecture/.style={-latex,Plum!80!black, dashed, line width=1.2pt, blur shadow={shadow blur steps=5, shadow xshift=0.3mm, shadow yshift=-0.3mm}}]
    \foreach \lab [count=\c, 
                   evaluate=\c as \ang using {18+72*\c}] 
    in {\text{\textbf{CP-indivisible}}, \text{increase of $d_\mathrm{tr}$\hspace{15mm}}, \text{increase of $V_\mathrm{Bloch}$}, \text{increase of $\mon$}, \text{}} {
       \node[bullet, blur shadow={shadow blur steps=5, shadow xshift=0.3mm, shadow yshift=-0.3mm}] (\c) at (\ang:20mm) {};
       \node at (\ang:24mm){$\lab$};
    }
    \node[bullet, blur shadow={shadow blur steps=5, shadow xshift=0.3mm, shadow yshift=-0.3mm}] (p) at (18+72*2.5:22mm) {};
    \node at (18+72*2.52:32mm){\textbf{P-indivisible}};
    \node at (18+72*4.85:21mm){\textbf{violation of DPI}};
    \draw[implication] (2)--(1);
    \draw[implication] (4)--(1);
    \draw[implication] (3)--(2);
    \draw[implication] (3)--(1);
    \draw[implication] (p)--(1);
    \draw[implication] (2)--(p);
    \draw[implication] (3)--(p);
    \draw[implication] (5)--(1);
    \draw[conjecture] (4)--(2);
    \draw[conjecture] (5)--(2);
    \end{tikzpicture}
    \caption{Relations between different definitions and witnesses of quantum non-Markovianity. A solid line from node (A) to (B) means that it is proven that non-Markovianity according to (A) implies non-Markovianity according to (B). A dashed line means that among our results obtained via random sampling of two-step dynamics any dynamics which turned out to be non-Markovian with respect to (A) was also non-Markovian with respect to (B). Nodes in boldface denote definitions of non-Markovianity while other nodes represent witnesses.}
    \label{fig:implications-conjectures}
\end{figure}

Yet, CP-indivisibility can also be witnessed in other ways: for instance, monitoring the entanglement between the system and an initially entangled ancilla, see Ref.~\cite{RivHuePle2010}. 
If for an entanglement monotone $\mon$, such as the entanglement or concurrence of formation, the dynamics $\Dyn=\dynamics{\cpt_1, \cpt_2}$ satisfies
\begin{align}
    \label{eq:rhp-entanglement}
    \mon\left[E_1\right] < \mon\left[E_2\right] 
\end{align}
with $E_i$ being the Choi states of the maps $\cpt_i$,
the dynamics is CP-indivisible. However, not every CP-indivisible dynamics leads to an increase of the entanglement, hence Eq.~\eqref{eq:rhp-entanglement} is a more restrictive and less sensitive criterion for non-Markovianity, compared to CP-indivisibility.

Moreover, there are also valid physical quantum dynamics that are not only CP-indivisible but also P-indivisible, meaning that $\cpt_{21}$ is not even a positive map. 
Indeed, P-indivisibility can be used as a stricter definition of non-Markovianity. However, positivity of $\cpt_{21}$ is comparably hard to check and thus in most cases only witnesses can be used. A very prominent one is given by an increase of the trace distance \cite{BreLaiPii2009}
\begin{align}
    \dtr(\rho, \sigma) = \frac{1}{2}\tr\sqrt{\left(\rho-\sigma\right)^\dagger \left(\rho-\sigma\right)}
\end{align}
between two initial states $\rho$ and $\sigma$ under the dynamics.
If there is at least one pair $(\rho, \sigma)$ of initial states such that $\dtr(\cpt_1(\rho), \cpt_1(\sigma)) < \dtr(\cpt_2(\rho), \cpt_2(\sigma))$, the dynamics is P-indivisible.
This witness has a direct interpretation in terms of memory effects: the trace distance is a measure of the distinguishability of two quantum states. If thus two states are hard to distinguish after the application of the first map $\cpt_1$ to both of them but the distinguishability increases for $\cpt_2$, this implies that there is a ``backflow of information'' into the system \cite{BreLaiPii2009}.
Note that there is a generalized trace-distance criterion, which, in principle, can be used to identify all P-indivisible dynamics \cite{kossakowskiQuantumStatisticalMechanics1972, wissmannGeneralizedTracedistanceMeasure2015, breuerColloquiumNonMarkovianDynamics2016b}. We will, however, stick to the originally formulated trace-distance witness of non-Markovianity, which is only sufficient but not necessary for P-indivisibility.

An even stricter geometric witness for P-indivisibility that is mainly used in the case of qubits ($d=2$) witnesses non-Markovianity by an increase of the volume $V_\mathrm{Bloch}$ enclosed by the Bloch sphere under the dynamics \cite{lorenzoGeometricalCharacterizationNonMarkovianity2013a}.
For a given map this volume can be computed in the Bloch-representation using Refs.~\cite{JevPusJenRud2014, MilJevJenWisRud2014}.
The Bloch-representation or Pauli-Liouville-representation $\Lambda$ of a map $\cpt$ is defined as
\begin{align}
    \Lambda_{ij} = \tr(\sigma_i \cpt\left[\sigma_j\right]),
\end{align}
where the Pauli matrices $\sigma_k = \{\id, \sigma_x, \sigma_y, \sigma_z\}$ form a basis of the operators on the Hilbert space.

A third approach to define non-Markovianity arises from information theoretic considerations and is based on correlations in terms of quantum mutual information (QMI) \cite{LuoFuSong2012DPI}. For a quantum state $\rho_{\Anc \Sys}$ with system $\Sys$ and ancilla $\Anc$ the QMI is defined as
\begin{align}
\label{eq:QMI}
I(\Anc;\Sys):=I(\rho_{\Anc \Sys})=S\left[\rho_\Anc\right] + S\left[\rho_\Sys\right] - S\left[\rho_{\Anc\Sys}\right]    
\end{align}
where $S=-\mathrm{tr}(\rho \log \rho)$ is the von Neumann entropy.
An increase of the QMI is a violation of the data-processing inequality (DPI) and witnesses a breakdown of a Markov chain representation of information flow, indicating memory effects \cite{LuoFuSong2012DPI}. Thus, if one observes
\begin{align}
    \label{eq:dpi}
    I(\Anc;\Sys)_1 < I(\Anc;\Sys)_2,
\end{align}
we say the dynamics is non-Markovian according to DPI-violation.

The definitions and witnesses of non-Markovianity provided are connected by a range of known (and two conjectured) implications -- for
a visual depiction of these interrelations see Fig.~\ref{fig:implications-conjectures}.

\subsection{Sampling Random Dynamics}
\label{sec:sampling-random-dynamics}

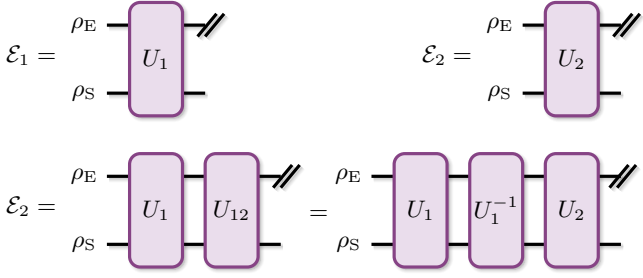
\begin{figure}
    \centering
    
    \begin{tikzpicture}
        \node at (-0.3,-0.6) {$\cpt_1=$};
        \unitarybox{1}{0}{$U_1$}{$\rho_\Env$}{$\rho_{\Sys}$}
        \traceout{2.05}{-0.15}
        \begin{scope}[xshift=5.5cm]
        \node at (-0.3,-0.6) {$\cpt_2=$};
        \unitarybox{1}{0}{$U_2$}{$\rho_\Env$}{$\rho_{\Sys}$}
        \traceout{2.05}{-0.15}
        \end{scope}
        \begin{scope}[yshift=-2cm]
        \node at (-0.3,-0.6) {$\cpt_2=$};
        \unitarybox{1}{0}{$U_{1}$}{$\rho_\Env$}{$\rho_{\Sys}$}
        \centerbox{2}{0}{$U_{12}$}
        \traceout{3.05}{-0.15}
        \begin{scope}[xshift=3.5cm]
        \node at (0,-0.7) {$=$};
        \unitarybox{1}{0}{$U_{1}$}{$\rho_\Env$}{$\rho_{\Sys}$}
        \centerbox{2}{0}{$U_{1}^{-1}$}
        \centerbox{3}{0}{$U_{2}$}
        \traceout{4}{-0.15}
        \end{scope}
        \end{scope}
    \end{tikzpicture}
    
    \caption{Physical motivation of the sampling strategy for a dynamics $\Dyn=\dynamics{\cpt_1, \cpt_2}$ used in this work. A randomly sampled first map $\cpt_1$ can be understood as the application of a random global system-environment unitary operation.
    While in reality, after having implemented $\cpt_1$ the map $\cpt_2$ is realized in terms of a second random unitary operation $U_{12}$ on system and environment, it suffices to generate a second map $\cpt_2$ independently. This can be seen by decomposing $U_{12}=U_2 \circ U_{1}^{-1}$, which implies that a randomly generated map corresponding to the consecutive random unitaries $U_1$ and $U_{21}$ contains the same information as an independently generated map corresponding to the unitary $U_2$. Note that the environment need not be of the same size as the system and can consist of many subenvironments such that the unitary operations $U_1$ and $U_{12}$ and hence also $U_1$ and $U_2$ may act trivially on some of those subenvironments.}
    \label{fig:sampling-scheme-interpretation}
\end{figure}

We would like to investigate the space of physically allowed two-step dynamics of the form $\Dyn = \dynamics{\cpt_1, \cpt_2}$. Our aim is to provide an estimate of the fraction of such dynamics which are non-Markovian with respect to those definitions and witnesses presented in the previous section.
For this purpose it is necessary to generate pairs of random quantum channels. In order to guarantee that this is done with respect to a proper measure, we follow the methods presented in Ref.~\cite{kukulskiGeneratingRandomQuantum2021a}.
This ensures that the maps are distributed equally according to the natural Lebesgue measure.

We will sample our maps in their Choi representation, see Eq.~\eqref{eq:def-choi}, so in the case of qubits we will create random two-qubit states.
These are obtained from the following three steps~\cite{kukulskiGeneratingRandomQuantum2021a}
\begin{enumerate}
    \item Generation of a random Wishart matrix $W=GG^\dagger$ of parameters $(d^2, M)$, where $G$ is a $d^2\times M$ random Ginibre matrix, $M$ is an integer and $d$ denotes the dimension of the Hilbert space of the quantum system.
    \item Tracing out the first half of $W$ defines $H=\tr_1 W$, which is, by construction, positive semidefinite.
    \item Computation of the normalized state\\ $E= \left(H^{-\frac{1}{2}} \otimes \id\right)W\left(H^{-\frac{1}{2}} \otimes \id\right)$.
\end{enumerate}
The obtained matrix $E$ satisfying $\tr_2 E=\id$ is the Choi state of a random CPT map $\cpt$. Its Kraus rank, the number of linearly independent rows in the matrix representation of the Choi state, is given by $k_\mathrm{max}=\min\{d^2, M\}$, so the integer $M$ can be used to control this rank $k_\mathrm{max}$.

We now combine two sampled maps given in their Choi representation to a dynamics $\Dyn=\dynamics{\cpt_1, \cpt_2}$. The space of all possible $\Dyn$ is the space we will be investigating with respect to definitions and witnesses of non-Markovianity provided earlier. The randomly generated two-step dynamics in this space can be seen as real-world instances of physical dynamics when considering an underlying system-environment picture, see Fig.~\ref{fig:sampling-scheme-interpretation}.

\section{Results}
\label{sec:results}
In the work we will focus on two-step qubit dynamics. To do so we randomly generate pairs of maps yielding $N=1.000.000$ dynamics of the form $\Dyn=\dynamics{\cpt_1, \cpt_2}$. In all cases the inverse $(\cpt_1)^{-1}$ existed and the corresponding intermediate maps $\cpt_{21}$ could thus be determined as explained earlier. These $\Dyn$ are then analyzed with respect to non-Markovianity according to the following five witnesses, see Sec.~\ref{sec:non-markovian-dynamics}
\begin{itemize}
    \item Non-Markovianity defined by CP-indivisibility
    \begin{enumerate}
        \item definition of CP-indivisibility \cite{RivHuePle2010}
        \item increase of an entanglement monotone \cite{RivHuePle2010}
    \end{enumerate}
    \item Non-Markovianity defined by P-indivisibility
    \begin{enumerate}
        \setcounter{enumi}{2}
        \item increase of the trace distance \cite{BreLaiPii2009}
        \item increase of the Bloch volume \cite{lorenzoGeometricalCharacterizationNonMarkovianity2013a}
    \end{enumerate}
    \item Non-Markovianity defined in terms of correlation dynamics (mutual information)
    \begin{enumerate}
        \setcounter{enumi}{4}
        \item violation of a data processing inequality \cite{LuoFuSong2012DPI}
    \end{enumerate}
\end{itemize}
The first and fifth witness are equivalent to the according definitions of non-Markovianity, the other three witnesses are only sufficient but do not necessarily detect non-Markovianity of that type for a dynamics $\Dyn$.

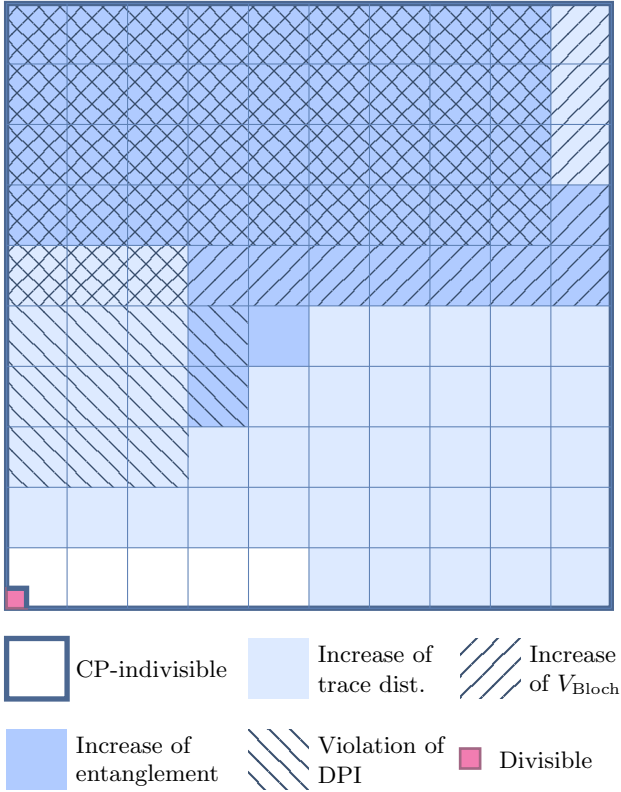
\begin{figure}
    \centering
    \begin{tikzpicture}[scale=0.8]

    \def\N{10}
    
    \fill[fill=mathematicablue!70] (0,0) rectangle (\N,\N);
    
    \fill[mathematicamediumblue!70] (3,3) rectangle (4,6);
    \fill[mathematicamediumblue!70] (4,4) rectangle (5,6);
    \fill[mathematicamediumblue!70] (5,5) rectangle (10,6);
    \fill[mathematicamediumblue!70] (0,6) rectangle (9,10);
    \fill[mathematicamediumblue!70] (9,6) rectangle (10,7);

    \fill[pattern={Lines[angle=45, line width=0.5pt, distance=2mm]}, pattern color=mathematicadarkblue!60!black] (0,5) rectangle (10,10);
    
    \fill[pattern={Lines[angle=-45, line width=0.5pt, distance=2mm]}, pattern color=mathematicadarkblue!60!black] (0,6) rectangle (9,10);
    \fill[pattern={Lines[angle=-45, line width=0.5pt, distance=2mm]}, pattern color=mathematicadarkblue!60!black] (0,2) rectangle (3,6);
    \fill[pattern={Lines[angle=-45, line width=0.5pt, distance=2mm]}, pattern color=mathematicadarkblue!60!black] (3,3) rectangle (4,5);
    
    \fill[white] (0,0) rectangle (5,1);
    \draw[line width=2pt, mathematicadarkblue!80!black] (0,0) rectangle (\N,\N);
    
    \foreach \i in {0,...,\N}
      \draw[thin, color=mathematicadarkblue] (\i,0) -- (\i,\N);
    \foreach \i in {0,...,\N}
      \draw[thin, color=mathematicadarkblue] (0,\i) -- (\N,\i);

    \filldraw[fill=VioletRed!80!white, draw= VioletRed!60!black, line width=1pt] (0,0) rectangle (0.33,0.33);
    \draw[line width=2pt, mathematicadarkblue!80!black] (0.33, 0) --(0.33, 0.33) --(0.0, 0.33);

    \filldraw[fill=white, line width=2pt, draw=mathematicadarkblue!80!black] (0,-1.5) rectangle ++(1,1);
    \node[] (a) at (1, -1) [right] {CP-indivisible};
    
    \fill[fill=mathematicablue!70] (4,-1.5) rectangle ++(1,1);
    \node[] (b) at (5, -0.75) [right] {Increase of};
    \node[] (b) at (5, -1.25) [right] {trace dist.};

    \fill[pattern={Lines[angle=45, line width=0.75pt, distance=2mm]}, pattern color=mathematicadarkblue!60!black] (7.5,-1.5) rectangle ++(1,1);
    \node[] (b) at (8.5, -0.75) [right] {Increase};
    \node[] (b) at (8.5, -1.25) [right] {of $V_\mathrm{Bloch}$};

    \fill[mathematicamediumblue!70] (0,-3) rectangle ++(1,1);
    \node[] (a) at (1, -2.25) [right] {Increase of};
    \node[] (a) at (1, -2.75) [right] {entanglement};
    
    \fill[pattern={Lines[angle=-45, line width=0.75pt, distance=2mm]}, pattern color=mathematicadarkblue!60!black] (4,-3) rectangle ++(1,1);
    \node[] (b) at (5, -2.25) [right] {Violation of};\node[] (b) at (5, -2.75) [right] {DPI};

    \filldraw[fill=VioletRed!80!white, draw=VioletRed!60!black, line width=1pt] (7.5,-2.65) rectangle ++(0.33,0.33);
    \node[] (b) at (8, -2.5) [right] {Divisible};

    \end{tikzpicture}
    \caption{Visualization of the fraction of $N=1.000.000$ randomly sampled two-step dynamics $\Dyn=\dynamics{\cpt_1, \cpt_2}$ which are non-Markovian according to different definitions and witnesses of non-Markovianity. In addition to the fractions for each of the criteria also their overlap with each other is depicted. See Tab.~\ref{tab:numerics-overlap} for details.}
    \label{fig:fraction-general-qubit-channel}
\end{figure}

The fractions of dynamics showing non-Markovianity according to each of those witnesses are shown in Fig.~\ref{fig:fraction-general-qubit-channel}. Inspired by the visualization in Ref.~\cite{Siu2020}, not only the relative values with respect to the total set of quantum dynamics are depicted, but also their overlaps.
More details and the exact percentages and uncertainties can be found in App.~\ref{sec:details-results}.

First we note that 99.90\% of the pairs $\Dyn$ turn out to correspond to CP-indivisible (non-Markovian) dynamics.
A remaining 0.10\% of the dynamics in the investigated space of two-step dynamics are CP-divisible (Markovian dynamics: pink lower left corner in Fig.~\ref{fig:fraction-general-qubit-channel}). Indeed, the data strongly suggests that the set of CP-divisible dynamics is very small, but has a nonzero measure within the set of all dynamics. Thus, only a tiny fraction of randomly chosen dynamics can be treated without incorporating memory effects. By contrast, 36\% of the dynamics are robustly non-Markovian: they are non-Markovian with respect to all of the five selected criteria.

While it is clear that P-indivisibility (and thus any criterion witnessing it) implies CP-indivisibility,
it is noticeable that although there is an overlap of the dynamics identified as non-Markovian with respect to the selected criteria, the sets do not form a strict hierarchy. Hence, each witness captures different aspects of the memory effects. This reflects the different physical motivations behind those concepts that complement each other.

By construction, the most sensitive criterion investigated here for witnessing non-Markovianity is indeed CP-indivisibility (99.90\%).
This is closely followed by the trace-distance witness (95.39\%) that detects P-indivisibility, appearing in the closely related BLP-non-Markovianity measure \cite{BreLaiPii2009}.

A crucial observation is that in our setup, the data processing inequality as well as the increase of the Bloch volume detect non-Markovianity in exactly 50\% of the investigated dynamics. This is for a simple reason: both witnesses compare the same quantity (QMI, Bloch volume), of two different maps. We can assume that the probability that the value for the first map $\cpt_1$ is exactly the same as as for the second map $\cpt_2$, is zero. Since we generate the random maps $\cpt_1$ and $\cpt_2$ independently of each other, the probability that they are sampled in the order $\Dyn=\dynamics{\cpt_1, \cpt_2}$ is the same as for the opposite order $\Dyn=\dynamics{\cpt_2, \cpt_1}$. Hence, in exactly half of the cases the two witnesses identify the dynamics as non-Markovian while in the other half they do not.

One could think that the same argument holds true for the witness concerning an increase of the entanglement. In that case, too, we compare the same quantity for two different maps.
However, unlike above, here the probability that two random maps lead to exactly the same entanglement value is nonzero. This is because there is a non-zero probability for a two-qubit state to be separable and thus return the value zero. In fact it is known that the fraction of separable states is exactly $8/33$ \cite{thanhhuongSeparabilityProbabilityTwoqubit2024, VolumeSeparableStates1998, MilzStrunz2015}.Taking this into account, one arrives at a theoretical fraction of of $(1-(8/33)^2)/2 = 47.06\%$ of all two-step dynamics to be non-Markovian with respect to an increase in entanglement which aligns with the result  $47.06\% \pm 0.06\%$ obtained from our Monte-Carlo results, see Tab.~\ref{tab:numerics-fractions}.
Sampling 1.000.000 dynamics, we found that dynamics showing an increase of an entanglement monotone and also those violating the DPI are strict subsets of dynamics showing an increase of the trace distance which hints at a possible implication. 
This may not be true in general, if for example the subset of dynamics violating a DPI but not featuring an increase of trace distance is of measure zero. We show in Fig.~\ref{fig:implications-conjectures} a visual depiction of known implications between witnesses of non-Markovianity and also the observed (conjectured) hierarchies from our random sampling.

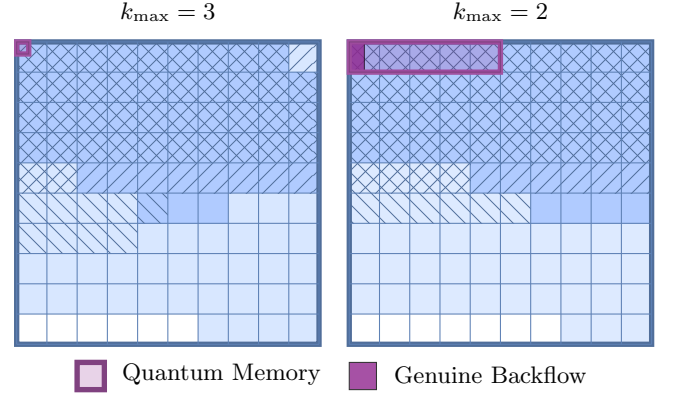
\begin{figure}
    \centering
    \begin{tikzpicture}[scale=0.4]
    \def\N{10}
    \node[] (a) at (5, 11) [] {$k_\mathrm{max}=3$};
    \fill[fill=mathematicablue!80] (0,0) rectangle (\N,\N);

    \fill[mathematicamediumblue!70] (2,5) rectangle (10,6);
    \fill[mathematicamediumblue!70] (4,4) rectangle (7,5);
    \fill[mathematicamediumblue!70] (9,6) rectangle (10,9);
    \fill[mathematicamediumblue!70] (0,6) rectangle (9,10);

    \fill[pattern={Lines[angle=45, line width=0.3pt, distance=1.5mm]}, pattern color=mathematicadarkblue!85!black] (0,5) rectangle (10,10);
    
    \fill[pattern={Lines[angle=-45, line width=0.3pt, distance=1.5mm]}, pattern color=mathematicadarkblue!85!black] (0,6) rectangle (9,10);
    \fill[pattern={Lines[angle=-45, line width=0.3pt, distance=1.5mm]}, pattern color=mathematicadarkblue!85!black] (9,6) rectangle (10,9);
    \fill[pattern={Lines[angle=-45, line width=0.3pt, distance=1.5mm]}, pattern color=mathematicadarkblue!85!black] (0,3) rectangle (4,5);
    \fill[pattern={Lines[angle=-45, line width=0.3pt, distance=1.5mm]}, pattern color=mathematicadarkblue!85!black] (4,4) rectangle (5,5);
    \fill[pattern={Lines[angle=-45, line width=0.3pt, distance=1.5mm]}, pattern color=mathematicadarkblue!85!black] (0,5) rectangle (2,6);
    
    \fill[white] (0,0) rectangle (6,1);
    \draw[line width=2pt, mathematicadarkblue!85!black] (0,0) rectangle (\N,\N);
    
    \foreach \i in {0,...,\N}
      \draw[thin, color=mathematicadarkblue] (\i,0) -- (\i,\N);
    \foreach \i in {0,...,\N}
      \draw[thin, color=mathematicadarkblue] (0,\i) -- (\N,\i);  

    \draw[line width=2pt, Plum!80] (0,9.6) rectangle (0.4,10);
    \draw[fill=Plum, opacity=0.2] (0,9.6) rectangle (0.4,10);

    \draw[line width=2pt, draw=Plum!80] (2,-1.5) rectangle ++(0.9,0.9);
    \draw[fill=Plum, line width=2pt, opacity=0.2] (2,-1.5) rectangle ++(0.9,0.9);
    \node[] (a) at (3.2, -1) [right] {Quantum Memory};

    \begin{scope}[xshift=11cm]

    \node[] (a) at (5, 11) [] {$k_\mathrm{max}=2$};
    
    \def\N{10}
    
    \fill[fill=mathematicablue!70] (0,0) rectangle (\N,\N);
    
    \fill[mathematicamediumblue!70] (0,6) rectangle (10,10);
    \fill[mathematicamediumblue!70] (4,5) rectangle (10,6);
    \fill[mathematicamediumblue!70] (6,4) rectangle (10,5);

    \fill[pattern={Lines[angle=45, line width=0.3pt, distance=1.5mm]}, pattern color=mathematicadarkblue!85!black] (0,5) rectangle (10,10);
    
    \fill[pattern={Lines[angle=-45, line width=0.3pt, distance=1.5mm]}, pattern color=mathematicadarkblue!85!black] (0,6) rectangle (10,10);
    \fill[pattern={Lines[angle=-45, line width=0.3pt, distance=1.5mm]}, pattern color=mathematicadarkblue!85!black] (0,4) rectangle (6,5);
    \fill[pattern={Lines[angle=-45, line width=0.3pt, distance=1.5mm]}, pattern color=mathematicadarkblue!85!black] (0,5) rectangle (4,6);
    
    \fill[white] (0,0) rectangle (7,1);
    \draw[line width=2pt, mathematicadarkblue!85!black] (0,0) rectangle (\N,\N);

    \foreach \i in {0,...,\N}
      \draw[thin, color=mathematicadarkblue] (\i,0) -- (\i,\N);
    \foreach \i in {0,...,\N}
      \draw[thin, color=mathematicadarkblue] (0,\i) -- (\N,\i);  
      
    \draw[fill=Plum, opacity=0.8] (0,9) rectangle (0.5,10);
    \draw[line width=2pt, Plum!80] (0,9) rectangle (5,10);
    \draw[fill=Plum, opacity=0.2] (0,9) rectangle (5,10);
    
    \draw[fill=Plum, opacity=0.8] (0,-1.5) rectangle ++(0.9,0.9);
    \node[] (a) at (1.2, -1) [right] {Genuine Backflow};
    
    \end{scope}
    \end{tikzpicture}
    \caption{Visualization of the fraction of randomly sampled two-step dynamics as in Fig.~\ref{fig:fraction-general-qubit-channel}, but for the maximal Kraus ranks $k_\mathrm{max}=3$ (left) and $k_\mathrm{max}=2$ (right). The matching between colors, patterns and criteria is the same as in Fig.~\ref{fig:fraction-general-qubit-channel}. Here, in addition, lower bounds for the fractions of dynamics with quantum memory and genuine backflow according to Sec.~\ref{sec:quantum-vs-classical} can be depicted.}
    \label{fig:influence-kraus-rank}
\end{figure}

\subsection{Influence of the Kraus rank of the channels}

Using the approach from Ref.~\cite{kukulskiGeneratingRandomQuantum2021a} in order to generate randomly sampled quantum channels, the maximal Kraus rank $k_\mathrm{max}$ of the channels can be fixed. For the investigations in Fig.~\ref{fig:fraction-general-qubit-channel} this was set to $k_\mathrm{max}=4$, which is the full rank and thus sufficient to describe any possible single-qubit dynamics.
We can now choose the cases $k_\mathrm{max}=2$ and $k_\mathrm{max}=3$ and observe how the corresponding fractions and intersections of the criteria differ from those in Fig.~\ref{fig:fraction-general-qubit-channel} of full rank. The results are depicted in Fig.~\ref{fig:influence-kraus-rank}.

It is noticeable that once the maximal Kraus rank $k_\mathrm{max}$ is set to be strictly below four, not a single dynamics is CP-divisible. In addition, for lower Kraus ranks less dynamics can be proven to be P-indivisible. The third criterion which changes significantly upon reducing $k_\mathrm{max}$ is the witness based on the increase of entanglement. This approaches 50\%, which reflects the fact that for channels of Kraus rank $k_\mathrm{max}=2$ there are only two eigenvalues contributing to the computation of the concurrence of formation. The probability that both have exactly the same value is zero and hence the argumentation as for the violation of the DPI or the increase of $V_\mathrm{Bloch}$ holds.

\subsection{Distinguishing classical from truly quantum memory effects}
\label{sec:quantum-vs-classical}

All approaches to witness non-Markovianity listed so far do not differentiate between memory effects which have a classical origin and those with a  truly quantum origin. This distinction, however, has recently gained much attention. Depending on the accessible information, concepts based on the process tensor \cite{MilEglTarThePleSmiHue2020, GiaCos2021, TarQuiMurMil2023}, and concepts based solely on dynamical maps \cite{BaeBeyStr2024, BusGanGosBadPanMohDasBer2025} have been put forward, for an overview see \cite{gangwarGenuineNonGenuineQuantum2026}.
As we concern ourselves with a dynamics given in terms of CPT maps and have no access to the process tensor, we will restrict ourselves to the approaches presented in Refs.~\cite{BaeBeyStr2024} and \cite{BusGanGosBadPanMohDasBer2025}. Note, however, that there is a close relation between the notions of classical or quantum memory in the approaches based on maps and process tensors~\cite{BaeLinStr2025}.

\subsubsection{Concepts verifying the quantumness of memory effects}
\label{sec:memory-effects}
Assume again that we are given a dynamics $\Dyn=\dynamics{\cpt_1, \cpt_2}$ such that we can compute their corresponding Choi states $E_1$ and $E_2$. 
One witness for the quantumness of the memory effects can be understood in the same spirit as the entanglement monotone criterion from Eq.~\eqref{eq:rhp-entanglement}, with some important modification. Instead of comparing the same quantity $\mon$ for two maps to witness non-Markovianity, quantum memory can be detected by checking whether \cite{BaeBeyStr2024}
\begin{align}
    \label{eq:qm-map}
    \mon^\sharp\left[E_1\right] < \mon \left[E_2\right],
\end{align}
where $\mon$ (as before) is the entanglement (or concurrence) of formation \cite{Wootters1997, Wootters1998}, yet $\mon^\sharp$ is the entanglement (or concurrence) of assistance \cite{DiVFucMabSmoThaUhl1999, LauVerEnk2002}.
While the former quantifies the minimal amount of average entanglement entropy required to form the quantum state $\rho$ from pure states, the latter characterizes that maximal possible value so that for any pure-state decomposition of $\rho$, the average entanglement entropy is less or equal to $\mon^\sharp\left[\rho\right]$. Thus, if the minimal attainable value of entanglement at an earlier time is exceeded at a later time, this is sufficient (but not necessary) to indicate that there is a ``backflow of \emph{truly quantum} information''. For more details on this classification and the criterion see App.~\ref{sec:review-concepts} or Ref.~\cite{BaeBeyStr2024}.

A different yet related approach \cite{BusGanGosBadPanMohDasBer2025} is based on the observation that memory effects in non-Markovian dynamics might be traced back to initial correlations with a dynamically inert auxiliary system ({\it non-causal revival}). By contrast, if no such explanation can be found, we observe {\it genuine backflow}. The distinction can be witnessed by quantum mutual information and the violation of the data processing inequality (DPI) Eq.~\eqref{eq:dpi}. A sufficient criterion for verifying genuine backflow is given by \cite{BusGanGosBadPanMohDasBer2025}
\begin{align}
    \label{eq:criterion-genuine}
    S\left[\tr_\Anc(E_1)\right] < \esq\left[E_2\right],
\end{align}
where $\esq$ is the squashed entanglement and the $E_i$ are the Choi states of the maps. For details on this criterion and the underlying concepts see App.~\ref{sec:review-concepts} or Ref.~\cite{BusGanGosBadPanMohDasBer2025}.

We will now investigate the fraction of non-Markovian dynamics resulting from genuine quantum memory (genuine backflow, resp.). First, let us note that the criterion from Eq.~\eqref{eq:qm-map} can directly be applied to qubit dynamics considered here. We simply choose the monotone $\mon$ to be the concurrence since in the qubit case there are explicit formulae to compute both, the concurrence of formation \cite{Wootters1998} and its dual quantity, the concurrence of assistance \cite{LauVerEnk2002}. Unfortunately,
the quantities appearing in criterion Eq.~\eqref{eq:criterion-genuine} cannot be obtained in a straightforward way since the squashed entanglement is computationally hard, even for qubits \cite{songLowerBoundsSquashed2009}.
In order to make a statement about dynamics featuring genuine backflow, we will thus first show important implications between the different witnesses, making it possible to estimate a lower bound also of the fraction of dynamics showing genuine backflow.

For this purpose, let us first consider the less tight criterion for quantum memory from Ref.~\cite{BaeBeyStr2025} for two Choi states $E_1$ and $E_2$. This is formulated in terms of entropic quantities and verifies the quantumness of the memory if
\begin{align}
    \label{eq:entropic-witness}
    S\left[\tr_\Anc(E_1)\right] < \max\left\{-S_{\Sys|\Anc}\!\left[E_2\right], -S_{\Anc|\Sys}\!\left[E_2\right]\right\},
\end{align}
where $S_{\Sys|\Anc}$ is the conditional quantum entropy
\begin{align}
    S_{\Sys|\Anc}\left[\rhoSA\right] :=S\left[\rhoSA\right] - S\left[\rho_\Sys\right].
\end{align}
If the entropic criterion Eq.~\eqref{eq:entropic-witness} witnesses quantum memory, necessarily also Eq.~\eqref{eq:qm-map} identifies the quantumness of the memory \cite{BaeBeyStr2025}.

Going further, we can easily show that for a given two-step dynamics $\Dyn=\dynamics{\cpt_1, \cpt_2}$ Eq.~\eqref{eq:entropic-witness} not only implies quantum memory (Eq.\eqref{eq:qm-map}), but also genuine backflow according to Eq.~\eqref{eq:criterion-genuine}.
This becomes immediately clear when considering the right-hand sides of those equations and noting that \cite{CarLie2012}
\begin{align}
    \max\left\{-S_{\Sys|\Anc}\!\left[E\right], -S_{\Anc|\Sys}\!\left[E\right]\right\}   \leq \esq\left[E\right],
\end{align}
for a bipartite state $E$. Since the left-hand sides of Eqs.~\eqref{eq:entropic-witness} and \eqref{eq:criterion-genuine} are already identical, the statement follows directly.

Note that if the monotone $\mon$ is chosen to be the entanglement entropy, we also have Eq.~\eqref{eq:criterion-genuine}$\Rightarrow$ Eq.~\eqref{eq:qm-map}. 
Hence, the criterion for genuine backflow also implies the presence of quantum memory. However, this does not necessarily imply that Eq.~\eqref{eq:qm-map} with $\mon$ being the {\it concurrence} can be used to witness it. This is due to the fact that while there is a function mapping the concurrence of formation to the entanglement of formation, no such relation between the concurrence of assistance and the entanglement of assistance is known \cite{DiVFucMabSmoThaUhl1999}.

Hence, if quantum memory according to Eq.~\eqref{eq:entropic-witness} has been identified, the dynamics necessarily also shows genuine backflow according to Eq.~\eqref{eq:criterion-genuine}. The fraction of dynamics identified by Eq.~\eqref{eq:entropic-witness} can thus serve as a lower bound for the fraction of dynamics with genuine backflow from Ref.~\cite{BusGanGosBadPanMohDasBer2025}.
In fact, although there seems to be a close relation between {\it quantum memory} \cite{BaeBeyStr2024} and {\it genuine backflow} \cite{BusGanGosBadPanMohDasBer2025}, these two concepts are different. While the concepts of classical and quantum memory are completely agnostic about the environmental embedding of the dynamics, genuine backflow and non-causal information revival depend on the particular choice of the environment. See App.~\ref{sec:comparison-qm-gb} for details and a proof that any mixed-unitary dynamics can be realized with non-causal information revival.

\subsubsection{Sampling Quantum Memory}
\begin{figure}
    \centering
    \includegraphics[width=1\linewidth]{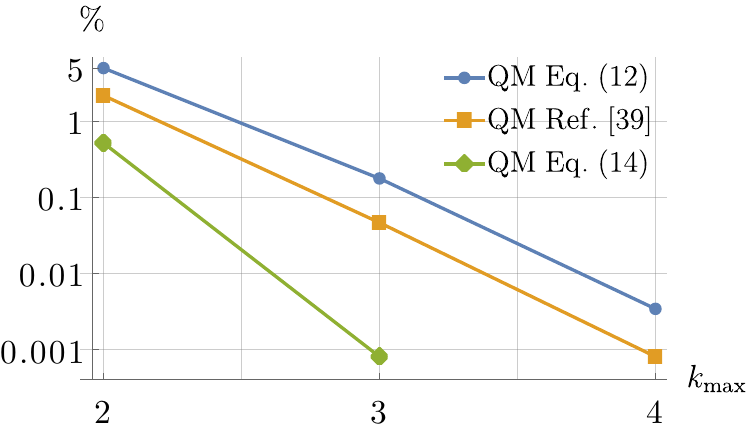}
    \caption{Dependence on the maximal Kraus rank $k_\mathrm{max}$ of the fraction of dynamics identified as having quantum memory. The original criterion for quantum memory (QM) refers to Eq.~\eqref{eq:qm-map}, the I-Concurrence criterion to the witness established in Ref.~\cite{BaeLinStr2025} and the entropic witness to Eq.~\eqref{eq:entropic-witness}. The latter one also serves as a computational tool to identify dynamics requiring genuine backflow in Eq.~\eqref{eq:criterion-genuine}. All these criteria provide lower bounds.}
    \label{fig:quantum-memory-kraus-rank}
\end{figure}

The absolute number of dynamics showing quantum memory according to Eq.~\eqref{eq:qm-map} is too low to be visible in the graphical depiction in Fig.~\ref{fig:fraction-general-qubit-channel}. However, once we consider lower maximal Kraus ranks, this changes, see Fig.~\ref{fig:influence-kraus-rank}. 
A more detailed depiction for the different witnesses of quantum memory can be seen in Fig.~\ref{fig:quantum-memory-kraus-rank}.
We observe that for the maximal Kraus rank $k_\mathrm{max}=2$ quantum memory is witnessed in almost 5\% of the investigated dynamics. Remarkably, the stricter entropic witness -- also identifying genuine backflow -- still captures 0.5\% of the randomly sampled dynamics. For a maximal Kraus rank $k_\mathrm{max}=4$, by contrast, the witness does not detect a single randomly generated dynamics.
Note that in addition to the two presented criteria, in Fig.~\ref{fig:quantum-memory-kraus-rank} we also show results for the sufficient witness from Ref.~\cite{BaeLinStr2025}, which can also be used in higher-dimensional systems and thus serves as a benchmark for further investigations.

An intuitive explanation for the observation that quantum memory becomes more relevant for lower Kraus ranks might be based on considering the realization of the dynamics in terms of a Stinespring dilation. Dynamics with low Kraus ranks can be realized with a smaller, and thus, more quantum environments.

\subsection{Mixed-unitary dynamics}
\label{sec:mixed-unitary}

An important subclass of real-world dynamics is given by mixed-unitary channels.
They have the form 
\begin{align}
    \label{eq:MU}
    \cpt_\text{MU}[\rho] = \sum_i^N p_i U_i \rho U_i^\dagger, && p_i \geq 0, && \sum_i p_i =1,
\end{align}
involving $N$ Kraus operators proportional to unitaries. In other words, such channels
can be expressed as a classical probabilistic mixture of unitary evolutions.
It is known that despite covering non-Markovian quantum dynamics, mixed-unitary dynamics can never display quantum memory \cite{BaeBeyStr2024, BaeBeyStrQMprecludesMU2026}. Furthermore, it can also be shown that any revival in mixed-unitary dynamics always admits a non-causal explanation, so genuine backflow is not necessary. For a proof see App.~\ref{sec:mu-ncir-cm}.
Therefore, an interesting question is whether restricting our analysis to mixed-unitary dynamics, do we find less memory effects in general, or do we simply remove instances involving quantum memory?

In order to generate a random mixed-unitary channel, we can again follow Ref.~\cite{kukulskiGeneratingRandomQuantum2021a}. In addition to the three steps from Sec.~\ref{sec:sampling-random-dynamics} we have to perform a fourth step. This is complementary to the third step and takes the form
\begin{align}
    \tilde{E}= \left(E^{-\frac{1}{2}} \otimes \id\right)W\left(E^{-\frac{1}{2}} \otimes \id\right),
\end{align}
where $E$ is the output of the normalization from step three. The result $\tilde{E}$ now becomes the input for the expression in the third step from Sec.~\ref{sec:sampling-random-dynamics} and one has to iterate those two steps until convergence is reached.

Although the naive expectation that for mixed-unitary dynamics $\mathcal{D}$ more instances are Markovian compared to the case of general two-step dynamics is true, the fraction of Markovian dynamics is still extremely low. Approximately a mere 0.4\% of the dynamics do not feature memory effects, see Tab.~\ref{tab:numerics-fractions}. The other fractions for the different criteria correspond more or less to those of arbitrary quantum channels. This shows for example that a random Choi state of a mixed-unitary map has the same probability to be separable as for general bipartite (Choi)-states, namely 8/33. 
The underlying reason is that the Choi state of a mixed-unitary map is a mixture of maximally entangled states \cite{DiVFucMabSmoThaUhl1999, AudSch2008}, which in turn implies that its reduced state is the identity. For a fixed reduced state, however, the fraction of separable states takes the same known value 8/33 as general bipartite qubit states \cite{MilzStrunz2015}. 
Nevertheless, differences arise when looking at the overlaps of the fractions, see Tab.~\ref{tab:numerics-overlap}.

\section{Conclusion}
\label{sec:conclusion}
In this work we investigate how often (non)-Markovianity occurs among randomly sampled time-discrete qubit dynamics.
For this purpose we choose five of the most prominent criteria and witnesses to identify memory effects in quantum dynamics. Those criteria are applied to time-discrete two-step qubit dynamics and their relative proportions with respect to the full space of those dynamics are estimated. 
The results show that in general a non-Markovian treatment of a generic qubit evolution is necessary and Markovian approaches such as the GKSL master equation might be too restrictive.
This is due to the fact that only 0.10\% of the dynamics are divisible and allow for a treatment with tools developed in the context of Markovian (CP-divisible) dynamics.
Roughly one third of the dynamics can be considered to be robustly non-Markovian, as memory effects are identified by all of the five criteria.
This fraction even increases when the maximally allowed Kraus rank is set to be strictly less than four.
We observe further that as soon as the Kraus rank is reduced, not a single dynamics is identified as being CP-divisible. This might imply that CP-divisible maps have measure zero in the space of all qubit dynamics with maximal Kraus rank $k_\mathrm{max}<4$.

Moreover, we identify a lower bound for the fraction of non-Markovian quantum dynamics for which memory effects are genuinly quantum and cannot be explained with classical memory. 
In order to also incorporate an estimation of the fraction of dynamics requiring a genuine backflow of information, we establish a connection to sufficient but not necessary criteria for quantum memory.
Although those criteria represent lower bounds only, the proportion of dynamics strictly requiring a genuinely quantum explanation for their occurrence seems to be very small.
This changes, however, when the maximal Kraus rank of the dynamics is reduced. While such a reduction also affects the fractions of the various non-Markovianity criteria, in proportion, the influence on the increase of the fraction of dynamics showing quantum memory or genuine backflow is much stronger.
Generically, dynamics with memory effects whose origin is truly quantum seem to be rather rare. Thus, quantum memory in non-Markovian dynamics is a valuable resource if required for quantum information processing tasks, and a profound theoretical understanding of such processes as well as precise engineering are crucial.

The results obtained in this article provide a basis for further investigations.
First, it may be possible to derive analytical expressions for the fractions of the different witnesses and also their overlaps. This could shed further light on the underlying physical motivations and interrelations of the different definitions, witnesses and criteria for non-Markovianity.

Furthermore, the fraction of dynamics which are non-Markovian with respect to yet other witnesses such as those from Refs.~\cite{LuWanSun2010, scandiQuantumFisherInformation2025, ChrKos2012} could be investigated as well. This could either provide new complementary witnesses or help in identifying strict hierarchies among the witnesses.
As a further next step the investigations could be extended to higher dimensions. Many of the criteria investigated are only computationally accessible for qubit dynamics, such as the increase of the volume of the Bloch sphere \cite{lorenzoGeometricalCharacterizationNonMarkovianity2013a} or an increase of entanglement with an ancilla \cite{RivHuePle2010}. Thus, one would need to resort to different criteria or use less sensitive witnesses providing a lower bound in order to obtain comparable results.
A direct comparison of the fractions for qubits with higher dimensional systems can, however, directly be obtained when considering CP-indivisibility, the violation of the DPI, the entropic witness and the witness from Ref.~\cite{BaeLinStr2025}.

\section*{Acknowledgements}
We thank Konstantin Beyer for helpful discussions and feedback on the manuscript.
C.\,B.~acknowledges support by the German Academic Scholarship Foundation.

\bibliography{literature_nm}

\appendix

\section{Details about the results}
\label{sec:details-results}

The visualizations in Figs.~\ref{fig:fraction-general-qubit-channel}, \ref{fig:influence-kraus-rank} and \ref{fig:quantum-memory-kraus-rank} are based on numerical evidence obtained by sampling qubit channels and computing the respective fractions. The processed data for the ratios for different criteria can be seen in Tab.~\ref{tab:numerics-fractions}, while the data for the overlaps is depicted in Tab.~\ref{tab:numerics-overlap}.

\begin{table}[t]
    \centering
    \begin{tabular}{|l|c|c|c|c|}
    \hline
        &  $k_\mathrm{max}=4$ & $k_\mathrm{max}=3$ & $k_\mathrm{max}=2$ & MU\\
         \hline
         number of dynamics &1.000.000 &500.000 &500.000 & 100.000\\
         \hline
       1 CP-indivisible  & 99.90 & 100.00 & 100.00 & 99.63\\ 
       2 P-indivisible  & 95.39& 94.34& 92.85 & 95.35\\
       3 Increase of $V_\mathrm{Bloch}$  & 50.06& 50.08  & 49.92 & 50.01 \\
       4 Increase of $\mon$  & 47.06 & 49.77 & 49.96 & 47.10\\
       5 Violation of DPI  &50.01 & 50.08 & 49.94 & 49.83\\
       \hline
       QM acc. to Eq.~\eqref{eq:qm-map} & $\approx 0$ & 0.18 & 4.97 & - \\
       QM acc. to Ref.~\cite{BaeLinStr2025} & $\approx 0$ & 0.05 & 2.16 & -\\
       QM acc. to Eq.~\eqref{eq:entropic-witness} & 0 & $\approx 0$& 0.51 & -\\
       \hline
       $\sigma_\mathrm{max}$ of mean & 0.06 & 0.09 & 0.08 &0.13 \\
       \hline
    \end{tabular}
    \caption{Fraction of dynamics identified as being non-Markovian with respect to the five criteria investigated in Sec.~\ref{sec:results} as well as those witnessing quantum memory (QM) from Sec.~\ref{sec:quantum-vs-classical} in \% rounded to two decimals. The sampling of two-step dynamics $\Dyn$ has been performed in the space of all dynamics as well as in the space of lower maximal Kraus ranks $k_\mathrm{max}$ and mixed-unitary (MU) channels. For mixed-unitary channels the quantum memory criteria have not been investigated since no mixed-unitary dynamics necessarily requires quantum memory.
    The maximal standard deviation of the obtained mean values is denoted by $\sigma_\mathrm{max}$.}
    \label{tab:numerics-fractions}
\end{table}

\begin{table}[]
    \centering
    \begin{tabular}{|ccccc||c|c|c|c|}
        \hline
         \rot{1 CP-indivisible} & \rot{2 P-indivisible} & \rot{3 Increase of $V_\mathrm{Bloch}$} & \rot{4 Increase of $\mon$} & \rot{5 Violation of DPI} & $k_\mathrm{max}=4$ & $k_\mathrm{max}=3$ & $k_\mathrm{max}=2$ & MU\\
         \hline
         & & & & & 0.10 & 0 & 0 & 0.37 \\
         \cm & & & & & 4.51 & 5.65 & 7.15 & 4.28 \\
         \cm & \cm & & & & 32.80 & 32.93 & 32.86 & 33.65 \\
         \cm & \cm & & & \cm & 9.03 & 7.99 & 6.51 & 9.25\\
         \cm & \cm & & \cm & \cm & 2.21 & 1.02 & 0 & 2.38 \\
         \cm & \cm & & \cm & & 1.28 & 2.32 & 3.57 & 0.05 \\
         \cm & \cm & \cm & \cm & & 7.51 & 7.83 & 6.49 & 7.78 \\
         \cm & \cm & \cm & & \cm & 2.71&2.48 &3.53 & 1.31 \\
         \cm & \cm & \cm & & & 3.78 & 1.18 & 0 & 4.04 \\
         \cm & \cm & \cm & \cm & \cm & 36.06 & 38.58&39.90 & 36.89 \\
         \hline
    \end{tabular}
    \caption{Fraction of dynamics witnessed as being non-Markovian with respect to certain combinations of criteria, depicted as the regions in Figs.~\ref{fig:fraction-general-qubit-channel} and \ref{fig:quantum-memory-kraus-rank} in \% rounded to one decimal. Differences to 100\% arise from rounding errors. 
    The maximal standard deviation of the mean value is $\sigma_\mathrm{max}=0.10$.}
    \label{tab:numerics-overlap}
\end{table}

\section{Comparison of the concepts quantum memory and genuine backflow}
\label{sec:comparison-qm-gb}

\subsection{Review of the concepts}
\label{sec:review-concepts}

A dynamics $\Dyn=\dynamics{\cpt_1, \cpt_2}$ can be realized with classical memory, if there exist Kraus operators $\{K_i\}$ and a suitable set of CPT maps $\Phi_i$ such that \cite{BaeBeyStr2024}
\begin{align}\label{eq:map_def}
        \cpt_{1}[\rho] = \sum_i K_i \rho K_i^\dagger, && \cpt_{2}[\rho] = \sum_i \Phi_i[K_i \rho K_i^\dagger].
\end{align}
Otherwise the dynamics is said to require quantum memory.
A non-Markovian dynamics with classical memory can be interpreted in terms of conditioned evolution based on measurement results $i$. The first map is realized by a generalized measurement with operators $K_i$ and the outcome $i$ of this measurement is recorded. In the second step -- conditioned on this outcome -- a CPT map $\Phi_i$ is realized. Summing up all possible outcomes yields the map $\cpt_2$. The crucial point why the memory can be considered classical is the fact that only the \emph{classical} measurement outcome $i$ has to be stored in a register and no quantum correlations need to be preserved in order to process from $\cpt_1$ to $\cpt_2$ (the CPT maps $\Phi_i$ are realized with a "fresh" environment).
In order to verify that a given dynamics only requires classical memory, at least one decomposition in terms of Kraus operators and CPT maps such as in Eq.~\eqref{eq:map_def} has to be found, which is, in general, a hard problem \cite{BaeBeyStr2024, YuOhsNguNim2025:p, BaeBeyStrQMprecludesMU2026}.
It is clear, however, that if the first map $\cpt_1$ is mixed-unitary, \emph{any} second map $\cpt_2$ in the dynamics can be realized using only classical memory \cite{BaeBeyStrQMprecludesMU2026}.

The concepts from Ref.~\cite{BusGanGosBadPanMohDasBer2025} treat the similar problem of whether revivals need a genuine backflow of information or whether there is, in principle, a noncausal explanation.
Using the concept of quantum mutual information from Eq.~\eqref{eq:QMI}, a revival satisfying the DPI Eq.~\eqref{eq:dpi} occurring in the dynamics $\Dyn=\dynamics{\cpt_1, \cpt_2}$ is said to be noncausal if a system-environment model including an inert environment extension $F$ exists such that the QMI of the tripartite state $\rho_{\Anc \Sys \F}$ decreases
\begin{align}
    I(\Anc;\Sys \F)_1 > I(\Anc;\Sys \F)_2,
\end{align}
otherwise the dynamics is said to require genuine backflow.
In order to understand why this characterizes dynamics with non-causal information revival, let us consider a global picture realizing the dynamics $\Dyn=\dynamics{\cpt_1, \cpt_2}$. This takes the form
\begin{align}
    \label{eq:system-environment-model}
    \rho_0 \mapsto U_{\Sys \Env} \rho_0 U_{\Sys \Env}^\dagger \mapsto V_{\Sys \Env} U_{\Sys \Env} \rho_0 U_{\Sys \Env}^\dagger V_{\Sys \Env}^\dagger,
\end{align}
where $U_{\Sys \Env}$ and $V_{\Sys \Env}$ are unitary operations on system $\Sys$ and environment $\Env$ and the initial state is given by
\begin{align}
    \label{eq:system-environment-model-state}
    \rho_0=\rho_{\Anc \Sys \Env \F}^0 = \Phi^+_{\Anc \Sys} \otimes \sigma_{\Env \F},
\end{align}
with $\Phi_{\Sys \Anc}^+$ being a maximally entangled Bell state of the system with an ancilla $\Anc$ and $\sigma_{\Env \F}$ denoting an initial state of the environment $\Env$ and an auxiliary environment extension $\F$. Both, $\Anc$ and $\F$ never directly interact with the system $\Sys$ and environment $\Env$, respectively.
The maps $\cpt_i$ can be extracted by tracing out additional degrees of freedom of the global states $\rho_{\Anc \Sys \Env \F}^i$, for $\cpt_1$ for instance this yields
\begin{align}
    \cpt_1\left[\rho_\Sys\right] = \tr_\mathrm{\Anc \Env \F} \left[ U_{\Sys \Env} \left(\Phi^+_{\Anc \Sys} \otimes \sigma_{\Env \F}\right) U_{\Sys \Env}^\dagger\right].
\end{align}

The central idea behind the distinction between noncausal information revivals and genuine backflow is to incorporate possible initial correlations of the environment with its extension $\F$. For revivals for which those correlations are sufficient to explain the violation of the DPI there is thus no backflow in the strict sense, leading to a noncausal explanation.

However, in order to verify that the revival in a given dynamics can be explained via noncausal information revival it is necessary to find an explicit system-environment model as in Eqs.~\eqref{eq:system-environment-model} and \eqref{eq:system-environment-model-state}. This requires in particular finding a suitable environment extension $\F$, which is, in general, also considered to be a hard problem \cite{BusGanGosBadPanMohDasBer2025}.
Nevertheless, if there is an $\F$ such that for the intermediate state $\rho_{\Anc \Sys \Env \F}^1 =  U_{\Sys \Env} \rho_0 U_{\Sys \Env}^\dagger$ it holds that the quantum conditional mutual information $I(\rho_{\Anc \Sys \Env \F}):=I(\Anc;\Env|\Sys \F)=I(\Anc;\Sys\F\Env)-I(\Anc;\Sys \F)$ is
\begin{align}
    \label{eq:qcmi-0}
    I(\Anc;\Env|\Sys \F)_1=0,
\end{align}
then \emph{any} revival occurring in the dynamics features a noncausal explanation.

Since there is a close relation between the sufficient criteria for quantum memory Eq.~\eqref{eq:qm-map} and genuine backflow Eq.~\eqref{eq:criterion-genuine}, it would be natural to assume some overlap of the concepts \emph{genuine backflow} and \emph{quantum memory}, as well as of \emph{noncausal information revival} and \emph{classical memory}.
As we will show, there are dynamics for which this is the case but we will also discuss fundamental differences between the two approaches.

\subsection{Examples where both concepts coincide}

We have seen that any dynamics with quantum memory according to Eq.~\eqref{eq:entropic-witness} necessarily also shows genuine backflow with respect to Eq.~\eqref{eq:criterion-genuine}. Hence, there is a class of dynamics having both properties. One example for such a dynamics is given by the dynamics $\Dyn=\dynamics{\cpt_{AD}, \id}$ where $\id$ is the identity channel and $\cpt_\mathrm{AD}$ is the full amplitude damping channel given in terms of the Kraus operators $K_1 = \ketbra{0}{1}$ and $K_2 = \ketbra{0}{0}$.
Evaluating the entropic witness from Eq.~\eqref{eq:entropic-witness} we find that quantum memory is strictly necessary for this dynamics. As Eq.~\eqref{eq:entropic-witness} implies Eq.~\eqref{eq:criterion-genuine} the dynamics also necessarily requires genuine backflow.

For an example showing classical memory and noncausal information revival, we can consider the dynamics $\Dyn=\dynamics{\cpt_\mathrm{D}, \id}$ with $\id$ being the identity map and $\cpt_\mathrm{D}$ the fully depolarizing qubit channel
\begin{align}
    \cpt_D\left[\rho\right] = \tr\left[\rho\right]\frac{\id_2}{2}.
\end{align}
Since $\cpt_\mathrm{D}$ is a mixed-unitary channel, any revival always only requires classical memory, see Ref.~\cite{BaeBeyStrQMprecludesMU2026}.
Let us now investigate noncausal information revival. In Ref.~\cite{BusGanGosBadPanMohDasBer2025} a global model has been proposed, using the three-qubit unitary
\begin{align}
    U_{SE} = \sum_{i=0}^3 \sigma_\Sys^i \otimes \ketbra{i}{i}_\Env
\end{align}
with $\sigma_i$ denoting the Pauli matrices.
We can now consider the initial state of the global dynamics to be $\Phi^+_{\Anc \Sys} \otimes \Phi_{\Env \F}$ where $\Phi_{\Env \F}$ is a maximally entangled four-qubit state.
Evaluating the condition from Eq.~\eqref{eq:qcmi-0} shows that any revival occurring after the intermediate global state $\rho_{\Anc \Sys \Env \F}^1$ can be explained in terms of a noncausal revival of information \cite{BusGanGosBadPanMohDasBer2025}. 
Hence, this dynamics $\Dyn=\dynamics{\cpt_\mathrm{D}, \id}$ is an example for which the concepts of noncausal information revival and classical memory coincide.

\subsection{Mixed-unitary dynamics can always be realized with non-causal information revival}
\label{sec:mu-ncir-cm}

It is known that the only subclass of dynamics $\Dyn=\dynamics{\cpt_1, \cpt_2}$ which show classical memory regardless of the second step are those where the first map $\cpt_1$ is mixed-unitary. It is thus reasonable to assume that if there is a connection to the concept of noncausal information revival, those dynamics will satisfy Eq.~\eqref{eq:qcmi-0}.
Now, in order to show formally that any revival occurring in the dynamics  where $\cpt_1$ is a mixed-unitary (MU) channel features a non-causal explanation, we will extend the example from Ref.~\cite{BusGanGosBadPanMohDasBer2025}, which we have already encountered in the previous section, to a general MU channel as in Eq.~\eqref{eq:MU}.

As our initial state we choose $\rho_{\Anc \Sys \Env \F}=\ketbra{\Psi_{\Anc \Sys \Env \F}}{\Psi_{\Anc \Sys \Env \F}}$ with
\begin{align}
    \ket{\Psi_{\Anc \Sys \Env \F}} = \ket{\psi_{\Anc \Sys}}\ket{\Phi_{\Env \F}},
\end{align}
where $\ket{\psi_{\Anc \Sys}}$ is a maximally entangled state of $d^2$ dimensions and 
\begin{align}
    \ket{\Phi_{\Env \F}} = \sum_{i=1}^{N} \sqrt{p_i} \ket{i_\Env}\ket{i_\F}.
\end{align}
The $p_i$ are the probabilities from the mixed-unitary channel in Eq.~\eqref{eq:MU} and both, the environment $\Env$ and its extension $\F$, are of dimension $N$.
The dynamics is realized by a unitary operation on system and environment given by
\begin{align}
    U_{\Sys \Env} = \sum_{i=1}^{N} U_i \otimes \ketbra{i_\Env}{i_\Env}.
\end{align}
We can now compute the quantum conditional mutual information of the evolved state 
\begin{align}
    \label{eq:rhoasef-evolved}
    \ket{\Psi'_{\Anc \Sys \Env \F}}= U_{\Sys \Env} \ket{\Psi_{\Anc \Sys \Env \F}},
\end{align}
which can be formulated as
\begin{align}
\label{eq:qcmi-entropies}
    I(\Anc;\Env|\Sys \F) = S(\Anc \Sys \F) + S(\Sys \Env \F) - S(\Sys \F) - S(\Anc \Sys \Env \F).
\end{align}

Tracing out the corresponding subsystems we find
\begin{align}
\label{eq:asf}
    \rho'_{\Anc \Sys \F} &= \sum_{i=1}^N p_i \left[(\id_\Anc \otimes U_i) \rho_{\Anc \Sys} (\id_\Anc \otimes U_i^\dagger)\right] \otimes \ketbra{i_\F}{i_\F},\\
    \label{eq:sf}
    \rho'_{\Sys \F} &= \sum_{i=1}^N p_i (U_i \rho_\Sys U_i^\dagger) \otimes \ketbra{i_\F}{i_F},\\
    \label{eq:sef}
    \rho'_{\Sys \Env \F} &= \sum_{k=1} q_k\ketbra{\Psi^k_{\Sys \Env \F}}{\Psi^k_{\Sys \Env \F}},
\end{align}
where in the last line we used the diagonal representation of the initial state $\rho_\Sys$ given in terms of probabilities $q_k$ and pure states $\ket{\psi^k_\Sys}$ leading to
\begin{align}
    \Psi^k_{\Sys \Env \F} = \sum_{i=1}^N \sqrt{p_i} U_i \ket{\psi^k_\Sys} \ket{ii_{\Env \F}}
\end{align}

Now the entropies of those states can be computed, as necessary for Eq.~\eqref{eq:qcmi-entropies}.
We first observe that $S(\Anc \Sys \Env \F)_1=0$ since the global state is pure.
As the quantum-classical state $\rho'_{\Anc \Sys \F}$ is diagonal in orthogonal projectors, its entropy is the Shannon entropy of the $\{p_i\}$ yielding $S(\Anc \Sys \F)_1=H(p)$. The entropy of the quantum-classical state $\rho'_{\Sys \F}$ is given by $S(\Sys \F)_1=H(p)+S(\rho'_\Sys)$ and for $\rho'(\Sys \Env \F)$ we find $S(\Sys \Env \F)_1=H(q)=S(\rho'_\Sys)$.
Using those entropies we directly see that according to Eq.~\eqref{eq:qcmi-entropies} $I(\Anc;\Env|\Sys \F)_1=0$.
Hence, any revival occurring after a mixed-unitary map $\cpt_1$ is always non-causal.

\subsection{Differences between the concepts}

According to Ref.~\cite{BusGanGosBadPanMohDasBer2025}, any revival that occurs in the presence of an initially pure environment $\Env$ can never show only noncausal information revival, genuine backflow is necessary.
Let us thus consider such a dynamics, which is given in terms of the two-qubit unitary
\begin{align}
    \label{eq:se-model-u}
    U_{\Sys \Env}(t) = \mathrm{exp}\left({-\frac{\iu t}{2} \sigma_z \otimes \sigma_x}\right).
\end{align}
We consider the evolution
\begin{align}
    \label{eq:se-model-evolution}
    \cpt_t\left[\rho_{\Sys}(0)\right]= \tr_\Env \left[U_{\Sys \Env}(t) \left(\rho_\Sys(0) \otimes \rho_\Env(0) \right) U_{\Sys \Env}(t)^\dagger\right],
\end{align}
where $\rho_\Env(0)=\ketbra{1}{1}$ is a pure state such that there is no initial correlation between $E$ and $F$ allowing for a noncausal explanation of possible revivals.
Such a revival according to the DPI can be obtained when considering the dynamics $\Dyn = \dynamics{\cpt_\frac{\pi}{2}, \cpt_\pi}$. Hence, this particular system-environment model of the dynamics requires genuine backflow.

Observing that $\cpt\left[\id\right]=\id$ we see that the dynamics is unital and since we concern ourselves with qubit dynamics this is equivalent to being mixed-unitary.
According to Ref.~\cite{BaeBeyStrQMprecludesMU2026}, the dynamics thus only requires classical memory.
Hence, the particular system-environment model in Eqs.~\eqref{eq:se-model-u} and \eqref{eq:se-model-evolution} shows genuine backflow but only requires classical memory.

However, one could now argue that several different system-environment models can, in principle, lead to the the same local dynamics $\cpt_t$ as in Eq.~\eqref{eq:se-model-evolution} and in order to exclude the possibility that the revival has a noncausal explanation one would have to check \emph{all} possible global models without upper limits on the dimension of the Hilbert spaces of $\Env$ and $\F$. Although optimizing over all possible environments is not in the scope of Ref.~\cite{BusGanGosBadPanMohDasBer2025}, let us nevertheless for a moment simply assume that such a model (whose initial state of the environment is necessarily mixed) exists.
Although the definition of classical memory from Ref.~\cite{BaeBeyStr2024} only asks whether for a certain observed dynamics on the system there is at least one global picture which can be explained with classical memory and does not discuss particular choices of the environment, let us now also include the question of when such a representation in terms of classical memory can be attained.

In order to realize the locally observed dynamics described by Eq.~\eqref{eq:se-model-evolution} with classical memory, it is crucial that the initial state of the environment is \emph{not} mixed. The initial state of the environment has to be a pure state in order to allow for the observed revival to be realized with classical memory, see Refs.~\cite{GreWer2003, TS2011EnvironmentAssistedCorrection}.

Coming back to the assumption that classical memory and noncausal information revival are closely related concepts, we observe that in this particular example the concepts are at odds: on the one hand, for noncausal information revival to occur, an initially mixed environment is necessary, whereas on the other hand it rules out a realization with classical memory.
We can hence conclude that the concepts of classical memory and noncausal information revival are not identical on a global level, for example once a certain initial environmental state is fixed.
This is a crucial observation: the concept of genuine backflow is not agnostic about the environment $\Env$ but strongly depends on its choice. A non-causal information revival is defined in terms of the existence of a suitable environment extension $\F$ and the environment $\Env$ itself plays no role in the definition, although it clearly influences the classification. We can thus only discuss a relation of quantum memory and genuine backflow or classical memory and non-causal information revival, respectively, on the local level of the system. This becomes even more relevant in experimental settings as usually there is no access to the environment.

\end{document}